\documentclass[a4paper]{article}

\usepackage{INTERSPEECH2020}
\usepackage{amsmath}
 \usepackage{booktabs}
 \usepackage{tabu}
 \usepackage{multirow}
 \usepackage{multicol}
 \usepackage{xcolor,colortbl}
 \usepackage{color,soul}
 
\newcolumntype{?}{!{\vrule width 1.1pt}}
\newcommand{\hbl}{\noalign{
\hrule height 1.1pt
}}
\newcommand{\pad}{1.3}
\title{Learning to Recognize Code-switched Speech Without Forgetting \\
Monolingual Speech Recognition}

\name{Sanket Shah\textsuperscript{$+$}, Basil Abraham\textsuperscript{$*$}, Gurunath Reddy M\textsuperscript{$+$},\\
Sunayana Sitaram\textsuperscript{$+$}, Vikas Joshi\textsuperscript{$*$}}
\address{\textsuperscript{$+$}Microsoft Research India,\\
\textsuperscript{$*$}Microsoft Corporation}
\email{\{t-sansha, basil.abraham, t-gumadh, sunayana.sitaram, vikas.joshi\}@microsoft.com}

\begin{document}

\maketitle
\begin{abstract}

Recently, there has been significant progress made in Automatic Speech Recognition (ASR) of code-switched speech, leading to gains in accuracy on code-switched datasets in many language pairs. Code-switched speech co-occurs with monolingual speech in one or both languages being mixed. In this work, we show that fine-tuning ASR models on code-switched speech harms performance on monolingual speech. We point out the need to optimize models for code-switching while also ensuring that monolingual performance is not sacrificed. Monolingual models may be trained on thousands of hours of speech which may not be available for re-training a new model. We propose using the Learning Without Forgetting (LWF) framework for code-switched ASR when we only have access to a monolingual model and do not have the data it was trained on. We show that it is possible to train models using this framework that perform well on both code-switched and monolingual test sets. In cases where we have access to monolingual training data as well, we propose regularization strategies for fine-tuning models for code-switching without sacrificing monolingual accuracy. We report improvements in Word Error Rate (WER) in monolingual and code-switched test sets compared to baselines that use pooled data and simple fine-tuning.

\end{abstract}

\section{Introduction}

Code-switching is the use of more than one language in a single conversation or utterance and is prevalent in multilingual communities all over the world. Automatic Speech Recognition of code-switched speech is challenging due to the lack of transcribed code-switched speech data, a larger set of words to recognize, confusions between similar-sounding words in both languages and the lack of code-switched text data for language modeling, among other reasons. However, significant progress has been made in increasing the accuracy of ASR for code-switched languages with advances in acoustic modeling, language modeling, decoding strategies and data augmentation \cite{sitaram2019survey}.

Code-switched speech co-occurs with monolingual speech, and it is imperative that ASR systems that perform well on code-switched speech also perform well on monolingual speech of one or both languages that are being mixed. Research on code-switched ASR and code-switched speech and language processing in general has focused on improving the accuracy of models on code-switched test sets. There has been very little focus so far on the effect of these techniques on monolingual speech recognition.

In this paper, we investigate the effect of standard techniques such as data pooling and fine-tuning on code-switched data on both code-switched and monolingual test sets and show that while these techniques show improvements on code-switched speech, there is a loss in performance on monolingual test sets. We hypothesize that this happens due to the phenomenon of ``catastrophic forgetting", in which the model forgets the distribution of monolingual speech in favor of code-switched speech. 

We experiment with three languages - Telugu, Tamil and Gujarati and their code-switched counterparts with English. We address two situations - the first being when we have access to a monolingual model, but not monolingual training data and want to adapt the model to recognize code-switched speech without sacrificing monolingual accuracy. For this scenario, we propose using the Learning Without Forgetting (LWF) technique \cite{Li2018LearningWF} to ensure that the model does not forget the distribution of monolingual speech. The second scenario we address is when we have access to monolingual data in addition to the monolingual model. We propose various fine-tuning and regularization strategies that improve the performance on both code-switched and monolingual test sets. We suggest that future papers in code-switched speech and language processing should also report results on monolingual test sets to ensure that models can generalize across both code-switched and monolingual data.

The organization of this paper is as follows: Section 2 relates our work to prior work. Section 3 describes our training and test datasets and the experimental setup followed by a discussion of the results. Section 4 concludes.
% What CS is, why it is challenging for ASR systems

% How current CS ASR approaches focus on higher accuracies on a CS test set - how real-world scenarios will always have monolingual speech in addition to CS - important to do well on both. Perform better on CS while not regressing (or even doing better) on monolingual.

% Contributions:

% Experiments to show that pooling/fine-tuning with CS does hurt mono performance - baseline experiment numbers on 3 CS datasets

% Approaches to fine-tuning without hurting mono performance 

\section{Relation to prior work}

In our work, we use CTC-based end-to-end (E2E) models for speech recognition \cite{Graves_CTC_init,E2E_comp_5,E2E_comp_6,Luo2018TowardsEC}. 
Authors in \cite{Luo2018TowardsEC} propose hybrid CTC-Attention based E2E architecture for Mandrian-English code-switched ASR and also explore different modeling units, decoding strategies and the presence of language identification for code-switched ASR. \cite{endtoendasr} and \cite{choudhury-etal-2017-curriculum} propose different fine-tuning strategies to fine-tune monolingual models to improve performance on code-switched speech and text respectively. \cite{Winata2020MetaTransferLF} proposes a method of meta-transfer learning in a low-resource setting by judiciously extracting information from high-resource monolingual datasets followed by fine-tuning. 

Multi-task learning approaches have been studied extensively for improving code-switched speech recognition. \cite{cs-multi-task} and \cite{Song2017AML} proposes an approach based on Deep Neural Networks via Multi-task Learning (MTL-DNN) for Mandarin-English conversational speech recognition  (primary task) and language identification (LID) (auxiliary task). Features extracted during transfer learning are not specialized for the new task and can often be improved by fine-tuning. In this work, we present strategies for fine-tuning that improve performance on code-switched and monolingual speech recognition.

\cite{Li2018LearningWF} proposes a novel technique called Learning Without Forgetting (LWF) that learn the parameters for new task without degrading the performance on the old task \cite{Goodfellow2014AnEI,catastrophicforgetting}. LWF also assumes the unavailability of the old task training data, while learning the new task. In this paper, we use the LWF based approach to improve performance on code-switched speech, while not degrading performance on monolingual speech. Like in \cite{Li2018LearningWF} we assume that we have access to only code-switched speech training data. LWF provides a more direct way of preserving the distribution of monolingual speech leading to improved performance for recognition on both monolingual as well code-switched speech. 

% Multi-task learning, transfer learning, fine-tuning and other related approaches  have been studied extensively for improving code-switched ASRs. \cite{Song2017AML}

% CS ASR (few relevant refs only), end-to-end CS ASR

\section{Experimental Setup}

\subsection{Data}

We carried out experiments for three languages - Tamil (TA), Telugu (TE) and Gujarati (GU) and their code-switched counterparts with English - Tamil-English, Telugu-English and Gujarati-English. Although all three languages were mixed with English, the type and extent of mixing was different. We used two types of speech data for training - conversational data as well as phrasal data, which is similar to read speech, while for testing, only phrasal data was used. A straightforward method to ensure that a model is not over-optimized for code-switching is to include monolingual data in development and test sets. We test our models on monolingual and code-switched data sets separately to ensure that models perform well on both. Table \ref{tab:traintestdata} describes the dataset size in hours. 

% \textbf{Sanket/Gurunath: this table needs to have the mono and CM dataset hours separately. I think there is no need for a separate Dev set column. So you can mention Train+Dev (MONO), Train+Dev (CS), Test(MONO), Test (CS). Try to use CS instead of CM consistently since code-switching is more standard in speech. Also add CMI values for all CS datasets including train}

\begin{table}[!ht]
\caption{Training and test data statistics}
\label{tab:traintestdata}
\centering
\renewcommand{\arraystretch}{\pad}
\resizebox{0.45\textwidth}{!}{
\begin{tabular}{?>{\columncolor[HTML]{EFEFEF}}l|l|l|l|l?}
\hbl
\textbf{} & \cellcolor[HTML]{EFEFEF}\textit{\textbf{Train + Dev}} & \cellcolor[HTML]{EFEFEF}\textit{\textbf{Train + Dev }}& \cellcolor[HTML]{EFEFEF}\textit{\textbf{Test}} &\cellcolor[HTML]{EFEFEF}\textit{\textbf{Test}} \\ 

\textbf{} & \cellcolor[HTML]{EFEFEF}\textit{\textbf{(MONO)}} & \cellcolor[HTML]{EFEFEF}\textit{\textbf{(CS) }}& \cellcolor[HTML]{EFEFEF}\textit{\textbf{(MONO) }} &\cellcolor[HTML]{EFEFEF}\textit{\textbf{(CS) }} \\ \hline
TA& 212 hrs & 177 hrs (CMI: 22.08) & 24 hrs  & 19 hrs (CMI: 17.07)\\ \hline
TE& 170 hrs    & 243 hrs (CMI: 23.85) & 27 hrs  & 28 hrs (CMI: 21.62)                  \\ \hline
GU& 241 hrs& 186 hrs (CMI: 18.91)& 15 hrs  & 8 hrs (CMI: 16.32)             \\ 
\hbl
\end{tabular}
}
 \end{table}

The Code Mixing Index \cite{gamback2014measuring} is a metric that measure the amount of code-switching in a corpus by using word frequencies. We measure the CMI of our code-switched train and test sets and report them in parentheses in Table \ref{tab:traintestdata}. The CMI of Telugu-English is the highest, while Gujarati-English is the lowest suggesting that Telugu-English is the most code-switched while Gujarati-English is the least code-switched among the languages under consideration. We see the same CMI trend in phrasal and conversational data, suggesting that it holds across speaking styles in these languages.

\begin{table}[!ht]
\caption{Baseline Word Error Rates (WER)}
\label{tab:baselines123}
\renewcommand{\arraystretch}{\pad}
\centering
\begin{tabular}{?>{\columncolor[HTML]{EFEFEF}}l|l|l|l?}
\hbl
\cellcolor[HTML]{EFEFEF}\textbf{Test Set}        & \cellcolor[HTML]{EFEFEF}\textit{\textbf{Exp1}} & \cellcolor[HTML]{EFEFEF}\textit{\textbf{Exp2}} & \cellcolor[HTML]{EFEFEF}\textit{\textbf{Exp3}} \\ \hline
TA-MONO & 50.09 & 70.20 & \textbf{48.47}\\\hline
TA-CS   & 67.62 & 63.70 & \textbf{55.93}\\ \hline
TE-MONO & 46.90 & 57.52 & \textbf{45.15} \\\hline
TE-CS   & 59.91 & 44.46 & \textbf{40.75}\\ \hline
GU-MONO & 41.99 & 54.83 & \textbf{40.96}\\ \hline
GU-CS   & 51.68 & 47.50 & \textbf{45.92}\\ 
\hbl
\end{tabular}
\end{table}

\subsection{Baseline experiments}

We denote our training monolingual datasets {($X_1^L$, $Y_1^L$),...,($X_n^L$, $Y_n^L$)} where $L$ $\in$ \{TE/TA/GU\}, code-switched datasets {($X_1^{CS}$, $Y_1^{CS}$),...,($X_n^{CS}$, $Y_n^{CS}$)} where $CS$ $\in$ \{TE-EN/TA-EN/GU-EN\}. The labels $Y$ are graphemes and the character set includes the union of English and the respective language's characters. 
Our baseline model consists of two Convolution Neural Network (CNN) layers followed by five bidirectional long-short term (BLSTM) layers of  1024 dimension. Further, the frame-wise posterior distribution $P(Y|X)$ is conditioned on the input X and calculated by applying a fully-connected layer and a softmax function.
\begin{equation}
P(Y|X) = \textrm{Softmax}(Linear(h))
\end{equation}
where $h$ is the hidden state from BLSTM. The model parameters are trained using Connectionist Temporal Classification (CTC) \cite{10.1145/1143844.1143891} criterion. We use SGD optimizer with 3e-4 learning rate. The model is trained for 40 epochs, with mini-batch size equal to 64 per GPU.
We experimented with three baselines, which we refer to as Exp1, Exp2 and Exp3.

\begin{itemize}
    \item \textbf{Exp1}: Monolingual-only baseline, consisting of models trained only on monolingual data 
    \item \textbf{Exp2}: Code-switched-only baseline, consisting of models trained only on code-switched data
    \item \textbf{Exp3}: Pooled model, consisting of models trained using all the data from Exp1 and Exp2
\end{itemize}

Table \ref{tab:baselines123} shows Word Error Rates (WER) of all three baselines on both monolingual and code-switched test sets. Exp3, which is the pooled model consisting of monolingual and code-switched data performs best on all test sets. Exp1 performs better on monolingual test sets than Exp2, and the reverse is true for code-switched test sets, as expected. The n-gram Language Model (LM), trained with training transcriptions, is used during decoding. We do not apply any text normalization during LM training, which could take care removing non-standard spellings, known spelling variations that would further reduce all the WERs. Cross-transcription and borrowed words lower the accuracy on code-switched test sets, which can be mitigated by using a modified version of the WER, such as poWER \cite{srivastava2018homophone}. However, we use the regular version of WER metric in this work to enable a fairer comparison with monolingual test sets.

\begin{table}[!ht]
\caption{WER$[\%]$ of pooled and fine-tuned models}
\label{tab:baselines345}
\renewcommand{\arraystretch}{\pad}
\centering
\resizebox{0.28\textwidth}{!}{
\begin{tabular}{?>{\columncolor[HTML]{EFEFEF}}l|l|l|l?}
\hbl
\cellcolor[HTML]{EFEFEF}\textbf{Test Set}        & \cellcolor[HTML]{EFEFEF}\textit{\textbf{Exp3}} & \cellcolor[HTML]{EFEFEF}\textit{\textbf{Exp4}} & \cellcolor[HTML]{EFEFEF}\textit{\textbf{Exp5}} \\ \hline
TA-MONO &\textbf{48.47}& 50.03&50.64\\\hline
TA-CS   & 55.93& 55.68 & \textbf{55.55}\\ \hline
TE-MONO & \textbf{45.15} &46.43 &46.03\\\hline
TE-CS   & 40.75& 44.26&\textbf{40.03} \\ \hline
GU-MONO & 40.96& 46.42&\textbf{38.90}\\ \hline
GU-CS   & 45.92& 45.47&\textbf{42.51}\\ \hbl
\end{tabular}
}
\end{table}

\subsection{Fine-tuning experiments}

\subsubsection{Fine-tuning baselines}

In addition to the above baselines, we implement two baselines by fine-tuning pre-trained models with code-switched data with a 10\% lower learning rate than the base model. 

\begin{itemize}
    \item \textbf{Exp4}: Fine-tuning the Exp1 (monolingual) model on code-switched data
    \item \textbf{Exp5}: Fine-tuning the Exp3 (pooled) model on code-switched data
\end{itemize}

Table \ref{tab:baselines345} shows WERs of the fine-tuned models compared to the best baseline, which is the pooled model. We see that fine-tuning the pooled model performs best on all code-switched test sets, while it degrades performance on monolingual Tamil and Telugu test sets. On Gujarati, the fine-tuned model performs the best, and from the CMI we know that Gujarati is the least code-switched of the three datasets. Fine-tuning the monolingual model with code-switched data (Exp4) degrades performance on all three monolingual test sets.

\begin{table}[!ht]
\caption{WER $[\%]$ of models fine-tuned with varying amounts of data (D)}
\renewcommand{\arraystretch}{\pad}
\label{tab:finetunedata}
\centering
\resizebox{0.45\textwidth}{!}{
\begin{tabular}{?>{\columncolor[HTML]{EFEFEF}}l|l|l|l|l|l?}
\hbl
\cellcolor[HTML]{EFEFEF}\textbf{Test Set}        & \cellcolor[HTML]{EFEFEF}\textit{\textbf{Exp3}} & \cellcolor[HTML]{EFEFEF}\textit{\textbf{Exp5}} & \cellcolor[HTML]{EFEFEF}\textit{\textbf{D=75\%}} & \cellcolor[HTML]{EFEFEF}\textit{\textbf{D=50\%}} & \cellcolor[HTML]{EFEFEF}\textit{\textbf{D=25\%}} \\ \hline
TA-MONO& \textbf{48.47}    & 50.64 & 49.35                     & 49.06             &49.00      \\\hline
TA-CS& 55.93         & 55.55 & 55.08                         & 55.06         & \textbf{54.92} \\\hline
TE-MONO    & 45.15         & 46.03 & 44.57                         & 45.10         & \textbf{44.85} \\\hline
TE-CS   & 40.75     & 40.03 & 39.65                         & 40.02         & \textbf{39.32} \\\hline
GU-MONO    & 40.96         & 38.90 & 38.60                         & 38.43         & \textbf{38.21}        \\ \hline
GU-CS   & 45.92     & 42.51         & 42.42                     & 42.49     & \textbf{42.07}\\ 
 \hbl
\end{tabular}
}
\end{table}

\subsubsection{Varying the amount of data for fine-tuning}

We conduct an additional experiment in which we fine-tune models with varying amounts of code-switched data. For each epoch during fine-tuning, we randomly sample D\% of the data from the total CS data available. Table \ref{tab:finetunedata} shows WER on fine-tuning the pooled model (Exp3) with 25\%, 50\% and 75\% of the available code-switched data. Exp5 is fine-tuned on all the data.

We see from Table \ref{tab:finetunedata} that fine-tuning the model with only 25\% of the available code-switched data works best for five of the six test sets. More importantly, we find that the loss in performance in monolingual test sets (1-2\% absolute WER) is mitigated by fine-tuning on less code-switched data.

\subsubsection{Fine-tuning with regularization}

To prevent the model from regressing on the monolingual WERs during fine-turning, we try to minimize the Kullback-Leibler (KL) divergence between the output distributions of the pre-trained model and the fine tuned model. The final loss function after adding the regularization term during fine-tuning is as follows:
\begin{equation}
L = (1-\alpha)L_{CTC} + \alpha D_{KL}(P_t||Q_t)
\end{equation}

\begin{equation}
L = L_{CTC} + \gamma D_{KL}(P_t||Q_t)
\end{equation}

where $D_{KL}$ is the KL divergence loss term, $P_t$ is the output distribution from the pre-trained  model, $Q_t$ is the output distribution of the fine tuned model, $\alpha$ is tunable parameter for balancing the weight regularization term and CTC loss, $\gamma$ is the scaling term. For scaled KLD, we use $\gamma$ equals to 100.\\

\begin{table}[!ht]
\caption{WER $[\%]$ of models fine-tuned with KLD loss. Bold indicates improvements over previously described models}
\label{tab:kldregularization}
\renewcommand{\arraystretch}{\pad}
\centering
\resizebox{0.45\textwidth}{!}{
\begin{tabular}{?>{\columncolor[HTML]{EFEFEF}}l|l|l|l|l|l|l?}
\hbl
\cellcolor[HTML]{EFEFEF}\textbf{Test Set}        & \cellcolor[HTML]{EFEFEF}\textit{\textbf{$\alpha$=0.1}} & \cellcolor[HTML]{EFEFEF}\textit{\textbf{$\alpha$=0.3}} & \cellcolor[HTML]{EFEFEF}\textit{\textbf{$\alpha$=0.5}} & \cellcolor[HTML]{EFEFEF}\textit{\textbf{$\alpha$=0.7}} & \cellcolor[HTML]{EFEFEF}\textit{\textbf{$\alpha$=0.9}} & 
\cellcolor[HTML]{EFEFEF}\textit{\textbf{Scaled KLD}}\\ \hline
TA-MONO            & 49.92                   & 49.41                   & 49.50                   & 49.44                   & 49.28        &\textbf{48.38}        \\ \hline
TA-CS        & 55.37                   & 54.91                   & 55.06                   & 54.95          & 55.07                   & \textbf{53.63}\\ \hline
TE-MONO            & 44.32                   & \textbf{44.18}          & 45.66                   & 45.69                   & 45.15                   & 44.33\\ \hline
TE-CS       & 39.79                   & 39.62          & 39.88                   & 39.77                   & 39.67                   &39.74\\ \hline
GU-MONO           &       38.56                  &        38.88                 &     39.25                    &      39.30                   &   39.06 & 39.07                      \\ \hline
GU-CS       &  43.39                       & 43.27               &      43.02                   &       42.91                  &   42.99 & 43.31                      \\  \hbl
\end{tabular}
}
\end{table}

Table \ref{tab:kldregularization} shows results of regularization with KLD loss and scaled KLD. We see improvements on Tamil and Telugu over the previous fine-tuning experiments.

Overall, we see improvements over pooling (Exp3) and simple fine-tuning of the pooled model (Exp4) of 0.7-1\% absolute WER for monolingual test sets and 0.5-2\% absolute WER for code-switched test sets. Fine-tuning with less data on the pooled model works best across almost all test sets. However we see a slight improvement for Tamil monolingual and 1.7\% absolute improvement over the Tamil code-switched test set with scaled KLD .

  \begin{figure}[!htb]
    \center{\includegraphics[scale=0.6]{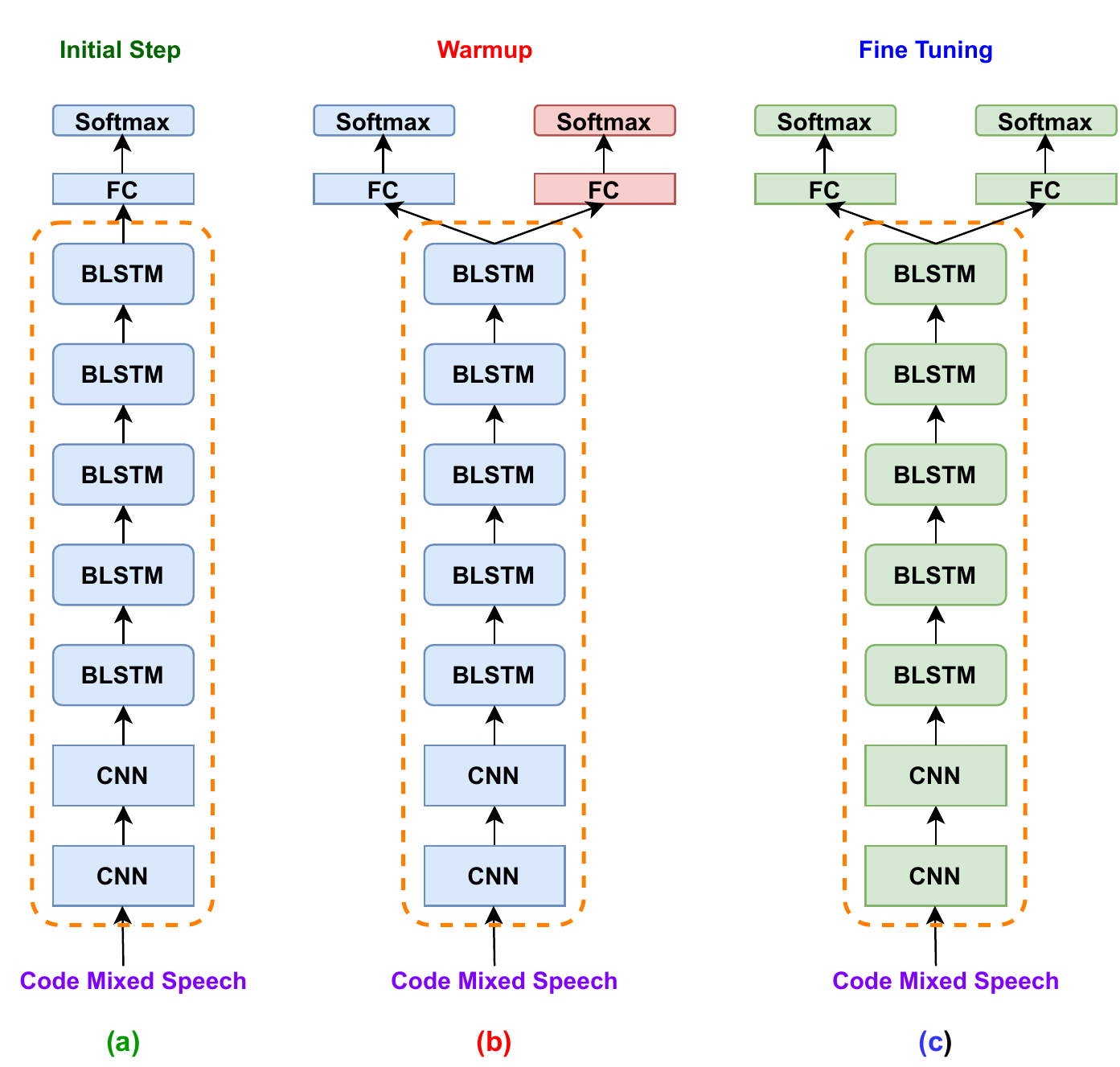}}
    \caption{Illustration of learning code-switched speech recognition without forgetting monolingual speech recognition. 
    (a) Pretrained monolingual model (old task) from which we obtain the labels of the old task, $Y_m$, by passing code-switched speech (new task data). (b) Code-switched task is added to the old monolingual task and the   model is trained with $Y_m$ and $Y_c$ as target labels by freezing old task parameters $\theta_m$, $\theta_s$ while updating new task parameters, $\theta_c$ with code-switched speech as input. (c) All parameters are unfreezed, and monolingual and code-switched speech recognition tasks are jointly trained. \label{fig:lwf-diagram}}
\end{figure}

\begin{table}[!ht]
\caption{Procedure for Learning without Forgetting}
\renewcommand{\arraystretch}{\pad}
\label{tab:lwfprocedure}
\centering
\begin{tabular}{?>{}l?}
\hbl

\textbf{Learning Code-Switched Speech Recognition Without}\\\textbf{Forgetting Monolingual Speech Recognition}\\ \\

\textbf{Start With:}\\

$\theta_s$: shared parameters\\
$\theta_m$: Task specific parameters of the monolingual \\ Speech Recognition task.\\
$X_c,Y_c$: Training data and ground truth for code-switched\\ Speech Recognition task.\\
\textbf{Initialize:}\\
$Y_m\xleftarrow[]{}$ pre-trained CTC-BLSTM$(X_c,\theta_s,\theta_m$)\\
$\theta_c\xleftarrow[]{}RandInit(\theta_c)$\\
\textbf{Train:}\\
Define $\hat Y_m$ $\equiv$ CTC-BLSTM$(X_c,\hat\theta_s,\hat\theta_m)$\\
Define $\hat Y_c$ $\equiv$ CTC-BLSTM$(X_c,\hat\theta_s,\hat\theta_m, \hat\theta_c)$\\
$\theta_s^*,\theta_m^*,\theta_c^* \xleftarrow[]{} \underset{\hat \theta_s, \hat \theta_m, \hat \theta_c}{\arg\max} \left( L_{CTC}(Y_m,\hat Y_m) + L_{CTC}(Y_c, \hat Y_c) \right)   $  \\ 

 \hbl
\end{tabular}
\end{table}

\subsection{LWF experiments}

In the previous experiments, we see that the best performing model is obtained when we fine-tune the pooled model with varying amounts of data or with KLD regularization. This requires access to all the monolingual data to build the pooled model. In cases where we do not have access to this data or do not want to rebuild a pooled model, we propose using the LWF framework. Thus, given a pre-trained monolingual model with shared parameters $\theta_s$, and task-specific parameters $\theta_m$ our aim is to add new task specific parameters $\theta_c$ for code-switched speech recognition and to learn parameters that work well for both monolingual as well code-switched speech recognition. The proposed application of LWF to code-switched ASR is shown in Fig.~\ref{fig:lwf-diagram}. The parameters of the CNN and BLSTM layers forms the shared parameters $\theta_s$ shown inside dotted box, while FC and softmax layers constitutes the task-specific layers ($\theta_m$ or $\theta_c$). The algorithm to train the proposed LWF is described in Table \ref{tab:lwfprocedure}.%The proposed LWF consists of three steps as shown in Fig.~\ref{fig:lwf-diagram}: a) We obtain the output for the new code-switched task from the old monolingual model which is defined by the parameters $\theta_s$ and $theta_m$ by passing code-mixed speech. b) We add new code-mixed task with parameters $\theta_c$ to the old monolingual model, initialized randomly.                  

%\textbf{Initial Step:} We first record transcripts $Y_o$ for each utterance in the code-switched training data. We initialize $\theta_n$ parameters, specific to the code-switched speech recognition task, randomly.

\textbf{Initial Step:} We first obtain the output $Y_m$ for the new code-switched task from the pre-trained monolingual model which is defined by the parameters $\theta_s$ and $\theta_m$ by passing code-switched speech shown in Fig.~\ref{fig:lwf-diagram}(a).

\textbf{Warm-up Step:} We add code-switched task with parameters $\theta_c$, initialized randomly to the pre-trained monolingual model. We freeze $\theta_s$ and $\theta_m$ and train $\theta_c$ for first 5 epochs. The warm-up step enhances further training of the model. The warm-up step is shown in Fig.~\ref{fig:lwf-diagram}(b).

\textbf{Fine Tuning Step}: When fine tuning, we unfreeze the parameters $\theta_s$ and $\theta_m$, and jointly train the model along with $\theta_c$ till convergence as shown in Fig.~\ref{fig:lwf-diagram}(c). In all steps, the model takes only code-switched speech as input.

Table \ref{tab:lwfresults} shows the comparison of standalone monolingual and code-switched models (Exp1 and Exp2) to the LWF model along with Exp4, which is the monolingual model fine-tuned with code-switched data. We also include the best fine-tuned model from the previous section as an upper-bound of what we can achieve if we did have access to the monolingual data.

The LWF model outperforms Exp1, Exp2 and Exp3 models across all test sets with the exception of Gujarati monolingual, for which the monolingual baseline performs slightly better. However, we see that the large drop in performance on Gujarati monolingual by the Exp4 model is mitigated to a large extent by LWF. Crucially, we see that the LWF models outperforms Exp4, which is the monolingual model with fine-tuning on code-switched data on all test sets. This indicates that the LWF framework can be used for building code-switched models without harming monolingual performance and without relying on having a pooled model for fine-tuning.

\begin{table}[!ht]
\caption{Word Error Rates (WER) of LWF experiment compared to baselines and best fine-tuned (FT) models}
\label{tab:lwfresults}
\renewcommand{\arraystretch}{\pad}

\centering
\resizebox{0.42\textwidth}{!}{
\begin{tabular}{?>{\columncolor[HTML]{EFEFEF}}l|l|l|l|l|l?}
\hbl
\cellcolor[HTML]{EFEFEF}\textbf{Test Set}        & \cellcolor[HTML]{EFEFEF}\textit{\textbf{Exp1}} & \cellcolor[HTML]{EFEFEF}\textit{\textbf{Exp2}} & \cellcolor[HTML]{EFEFEF}\textit{\textbf{Exp4}} & \cellcolor[HTML]{EFEFEF}\textit{\textbf{LWF}} &\cellcolor[HTML]{EFEFEF}\textit{\textbf{Best FT}}\\ \hline
TA-MONO &50.09& 70.20&50.03&\textbf{48.90} & 48.38\\\hline
TA-CS   & 67.62& 63.70 &55.68&\textbf{55.65}& 54.63\\ \hline
TE-MONO & 46.90 &57.52 &46.43&\textbf{45.70}& 44.18\\\hline
TE-CS   & 59.91&44.46&44.26&\textbf{43.60}&39.32\\ \hline
GU-MONO & 41.99& 54.83&46.42&\textbf{42.30}&38.21\\ \hline
GU-CS   & 51.68& 47.5&45.47&\textbf{44.20}&42.07\\\hbl
\end{tabular}
}
\end{table}

\section{Conclusion}

In this paper, we show that fine-tuning models for code-switching can lead to a drop in performance in monolingual models. Specifically, we show that a monolingual model that is fine-tuned on code-switched data improves on code-switched test sets, but degrades on monolingual test sets across ASR systems in three languages. A pooled model performs well on both monolingual and code-switched test sets, but fine-tuning this model with less code-switched data and regularization leads to best performance. 

Building and Fine-tuning a pooled model relies on the availability of monolingual data that the original ASR was trained on. This may not always be feasible given that monolingual ASRs can be trained on thousands of hours of data. To address this issue, we propose using the Learning Without Forgetting (LWF) framework to build code-switched models, without sacrificing monolingual accuracy. Our experiments show that models built using the LWF framework outperform monolingual models fine-tuned with code-switched data. We also show that the loss in performance on monolingual datasets is mitigated by using the LWF technique. More generally, the LWF framework can be used to adapt speech recognition models to specific domains without forgetting the distribution of the original data they were trained on.

One limitation of our experiments is that the amount of monolingual and code-switched data used for training are of the same order of magnitude, while this is not usually the case in real-world systems due to the lack of available code-switched data. In future work, we plan to replicate these experiments on monolingual models trained with hundreds of hours of data and fine-tuned with much less code-switched data.

As future work, we plan to explore adding BLSTM layers as task specific layers. We plan to investigate adding adversarial training procedure to make the shared layer parameters to be agnostic to the specific task while encouraging the model to learn discriminative parameters at the task specific layers. 

Code-switching is a special case of domain adaptation, because code-switched speech and text usually co-occur with monolingual speech and text. We suggest that future work in code-switched speech and text should also investigate the effects of models tuned for code-switching on monolingual test sets.

\bibliographystyle{IEEEtran}

\bibliography{mybib}

\end{document}